\documentclass[aps,prd,twocolumn,showpacs,nofootinbib]{revtex4-1}

\usepackage{amsfonts,color,graphicx,epsfig,amsmath,multirow}

\usepackage[colorlinks=true,linkcolor=blue,citecolor=blue, urlcolor=blue]{hyperref}

\begin{document}

\title{Photoproduction of the heavy quarkonium at the ILC}

\author{Gu Chen}
\author{Xing-Gang Wu}
\email{email:wuxg@cqu.edu.cn}
\author{Hai-Bing Fu}
\author{Hua-Yong Han}
\author{Zhan Sun}

\address{Department of Physics, Chongqing University, Chongqing 401331, P.R. China}

\date{\today}

\begin{abstract}

We study the photoproduction of the heavy quarkonium at the future International Linear Collider (ILC) within the nonrelativistic QCD theory. We focus on the production channel via the subprocess $\gamma\gamma \to |[Q\bar{Q'}]_{\bf 1}(n)\rangle+Q'+\bar{Q}$, where $Q$ and $Q'$ stand for heavy $c$- or $b$-quark, respectively. $|[Q\bar{Q'}]_{\bf 1}(n)\rangle$ stands for color-singlet $S$-wave quarkonium, i.e., $\eta_{c}(|[c\bar{c}]_{\bf 1}(^1S_0)\rangle)$, $J/\psi(|[c\bar{c}]_{\bf 1}(^3S_1)\rangle)$, $B_{c}(|[c\bar{b}]_{\bf 1}(^1S_0)\rangle)$, $B^*_{c}(|[c\bar{b}]_{\bf 1}(^3S_1)\rangle)$, $\eta_{b}(|[b\bar{b}]_{\bf 1}(^1S_0)\rangle)$, and $\Upsilon(|[b\bar{b}]_{\bf 1}(^3S_1)\rangle)$, respectively. To improve the calculation efficiency, we adopt the improved helicity amplitude approach to deal with the difficulty of calculating the expressions for the yields when the quark masses cannot be neglected. Total and differential photoproduction cross sections, together with their uncertainties, have been presented. It is noted that sizable amount of $|c\bar{c}\rangle$-charmonium and $|c\bar{b}\rangle$-quarkonium events can be generated at the ILC. More specifically, we predict $(2.8^{+1.0}_{-0.7})\times 10^{6}$ $\eta_c$, $(5.4^{+1.9}_{-1.3})\times 10^{6}$ $J/\psi$, $(8.3^{+2.2}_{-1.8})\times 10^{4}$ $B_c$, $(4.3^{+1.1}_{-0.9})\times 10^{5}$ $B_c^*$, $(9.0^{+1.7}_{-1.4})\times 10^{3}$ $\eta_b$, and $(1.6\pm0.3)\times 10^{4}$ $\Upsilon$ events to be generated in one operation year at the ILC under the condition of $\sqrt{S}=500$ GeV and ${\cal L}\simeq 10^{36}$cm$^{-2}$s$^{-1}$.

\end{abstract}

\pacs{13.66.Bc, 12.38.Bx, 12.39.Jh, 14.40.Pq}

\maketitle

\section{Introduction}

The International Linear Collider (ILC)~\cite{ILC1,ILC2} has been proposed and regarded as the next generation of the $e^+e^-$ collider. It is designed to run at a rather high center-of-mass energy from several hundred GeV to TeV together with a high luminosity up to ${\cal L}\simeq 10^{34-36}{\rm cm}^{-2}{\rm s}^{-1}$. At the high energy $e^+e^-$ collider, the photon beam can be generated by bremsstrahlung and be described by the Weiz$\ddot{a¡§}$cker-Williams approximation~\cite{WWA}. The laser backscattering (LBS) from the incident electron and positron beams leads to high luminosity photon beams, i.e., the LBS photons are hard enough and carry a large fraction of energy of the lepton beams. The density function of the incident photons can be found in Ref.~\cite{ydenfun}. In the literature, the $J/\psi$ photoproduction in $e^+e^-$ at the LEP II energy has been estimated within the color-singlet model~\cite{CSM} by two groups~\cite{yyJpsicc,yyJpsicc1}. Their results indicate that large production rates for $J/\psi$ in the direct photon collision. In view of a higher collision energy at the ILC, it is natural to expect that the ILC shall also provide an important platform for studying the heavy quarkonium properties. As the main purpose of the present paper, we shall make a detailed study on the photoproduction of the $S$-wave heavy quarkonium at the ILC.

\begin{figure}[htb]
\includegraphics[width=0.45\textwidth]{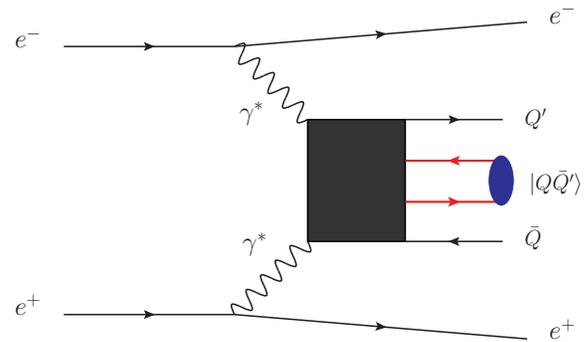}
\caption{The schematic Feynman diagram for the photoproduction of the $S$-wave heavy quarkonium in $e^+e^-$ scattering via the subprocess $\gamma\gamma \to |Q\bar{Q'}\rangle+Q'+\bar{Q}$, where $Q$ and $Q'$ stand for heavy $c$- or $b$-quark, respectively. The black box stands for the hard interaction kernel. } \label{schematic}
\end{figure}

The leading-order color-singlet heavy quarkonium photoproduction via the photon-photon collision based on the $e^+e^-$ collider can be schematically described by a diagram as shown in Fig.~\ref{schematic} \footnote{It is noted that the $2\to1$ subprocess $\gamma\gamma\to|Q\bar{Q'}\rangle$ provides dominant total cross-section for the quarkonium in $^{1}S_{0}$ state, however it has no phase-space distributions. In the present paper, we shall not take this special case into consideration.}. To deal with the production cross section, one needs the squared amplitudes, which are usually derived by applying the conventional trace technique, in which the squared amplitudes are first transformed into a trace form and then calculated. As will be shown in the next section, there are in total twenty Feynman diagrams for the subprocess $\gamma\gamma \to |Q\bar{Q'}\rangle+Q'+\bar{Q}$, where $Q$ and $Q'$ stand for heavy $c$- or $b$-quark, respectively. All the quark lines of the subprocess are massive, thus the results for its squared amplitudes are much too complex and lengthy. One important way to solve this is to deal with the process directly at the amplitude level. The helicity amplitude approach suggested by Refs.~\cite{helicity0,helicity} can be adopted for such purpose. Under the helicity amplitude approach, all the amplitudes are expressed in terms of helicity amplitudes, which are constant complex numbers and are immediately calculated. In its original version, the helicity amplitude approach has been designed to deal with the massless cases. Our present subprocesses contain non-Abelian gluons and massive fermions, thus an improved version has to be introduced. Several approaches for such purpose have been suggested in the literature, e.g., Refs.~\cite{helicity1,helicity2}. In the present paper, we shall adopt the way suggested by Ref.~\cite{helicity1} to do the calculation. The key point of this suggestion is to convert the problem into an equivalent `massless' one and to extend the `symmetries' as much as possible such that to achieve the most simplified amplitude.

The remaining parts of the paper are organized as follows. In Sec.II, we present the formulation for dealing with the subprocess $\gamma\gamma \to |Q\bar{Q'}\rangle+Q'+\bar{Q}$, where the improved helicity amplitude approach is adopted to simplify the hard scattering amplitude. In Sec.III, we give the numerical results. Sec.IV is reserved for a summary.

\section{Calculation technology}

\subsection{Differential cross section}

The production rates of the heavy quarkonium can be factorized into the short-distance and the long-distance parts within the framework of the nonrelativistic quantum chromodynamics (NRQCD)~\cite{nrqcd}. The short-distance coefficients can be calculated perturbatively. The non-perpurbative but universal long-distance matrix elements can be extracted from experimental measurements. The color-singlet matrix elements can be related with the wavefunction at the zero and be computed with certain potential models. Within the NRQCD framework, the differential cross section is formulated as
\begin{widetext}
\begin{equation}
d\sigma = \int dx_1 dx_2 f_{\gamma}(x_1) f_{\gamma}(x_2) d\hat\sigma(\gamma\gamma \rightarrow |(Q\bar{Q'})_{\bf (1)}[n]\rangle + \bar{Q} + Q') \langle{\cal O}^H(n) \rangle\;, \label{corsec}
\end{equation}
\end{widetext}
where $\langle{\cal O}^H(n) \rangle$ is the long-distance matrix element with $n$ standing for the intermediate $(Q\bar{Q'})$-pair state and $H$ being the final quarkonium state. Here, we will not consider the spin-flip effect between the $(Q\bar{Q'})$-pair and the quarkonium. It may provide sizable contribution for the polarized cross sections, which however is model dependent. A detailed discussion on the spin-flip effect for the hadronic $J/\psi$ production can be found in Ref.\cite{flip}. The subscript ${\bf (1)}$ means the intermediate $(Q\bar{Q'})$-pair is in color-singlet state. In this paper, we shall concentrate on the color-singlet $S$-wave heavy quarkonium production, i.e., $H=\eta_c$, $J/\psi$, $B_c$, $B_c^*$, $\eta_b$, and $\Upsilon$, respectively. Both the productions for the heavy quarkonium in higher Fock states, such as the color-singlet $P$-wave states and the color-octet $S$-wave states and etc., and the productions at the higher perturbative orders are much more involved, which are in progress~\cite{chenwu}. $f_{\gamma}(x)$ is the density function of the incident photons~\cite{ydenfun}
\begin{eqnarray}
f_{\gamma}(x) &=& \frac{1}{N} \left[1-x+\frac{1}{1-x}-4r(1-r)\right] ,
\end{eqnarray}
where $r=x/[x_m(1-x)]$ and the normalization factor
\begin{eqnarray}
N &=& \left(1-\frac{4}{x_m}-\frac{8}{x_m^2}\right)\log \chi +\frac{1}{2}+\frac{8}{x_m}-\frac{1}{2\chi^2}.
\end{eqnarray}
Here, $\chi=1+x_m$ and $x_m = \frac{4E_e E_l}{m_e^2} \cos^2\frac{\theta}{2}\simeq4.83$~\cite{lbs} with $E_e$ and $E_l$ being the energies of the incident electron and laser beams, respectively, and $\theta$ is the angle between those two beams. The energy of the LBS photon is restricted by
\begin{eqnarray}
0 \leq x \leq \frac{x_m}{1+x_m}=0.83.
\end{eqnarray}

The $2\to3$ short-distance differential cross section $d\hat\sigma$ can be written as
\begin{eqnarray}\label{hardc}
&& d\hat\sigma(\gamma\gamma \rightarrow |Q\bar{Q'}[n]\rangle + \bar{Q} + Q') = \frac{1}{2 x_1 x_2 S} \overline{\sum}  |{\cal M}|^{2} d\Phi_3 \;.
\end{eqnarray}
where $\sqrt{S}$ is the collision energy of the $e^+e^-$ collider, $\overline{\sum}$ means we need to average over the spin states
of the incident photons and to sum over the color and spin of
all final particles. $d\Phi_3$ is the three-body phase space,
\begin{equation}
d{\Phi _3} = {(2\pi )^4} {\delta ^4}({k_1} + {k_2} - \sum\limits_f^3 {{q_f}}) \prod\limits_{f = 1}^3 {\frac{d\vec{q}_{f}} {(2\pi)^3 2q_f^0}} .
\end{equation}

\begin{figure*}
\includegraphics[width=0.9\textwidth]{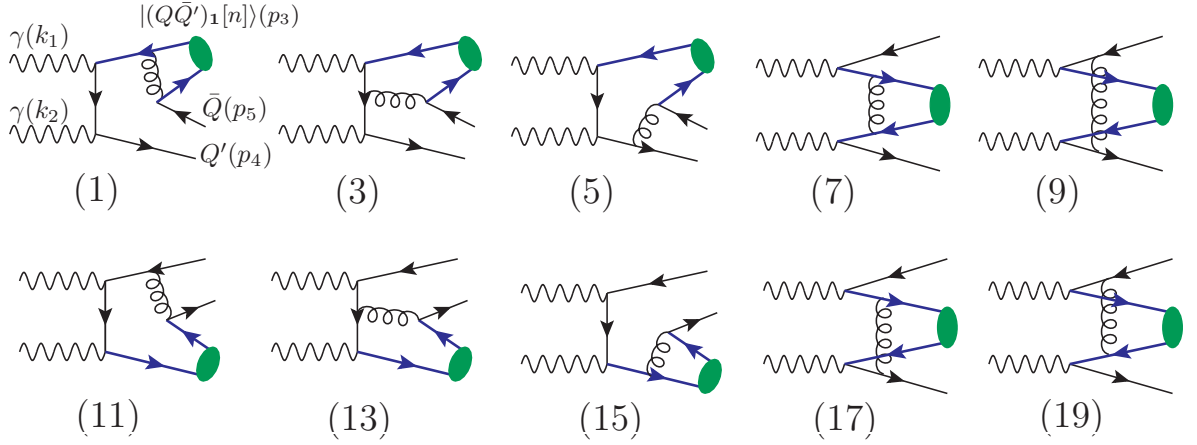}
\caption{Typical Feynman diagrams for the subprocess $\gamma(k_{1})\gamma(k_{2}) \to |(Q\bar{Q'})_{\bf 1}[n]\rangle(p_{3})+ Q'(p_4)+\bar{Q}(p_5)$. The remaining ten diagrams can be obtained by exchanging the position of the incident photons.} \label{YYFeny}
\end{figure*}

We present ten typical Feynman diagrams for the subprocess $\gamma(k_{1})\gamma(k_{2}) \to |(Q\bar{Q'})_{\bf 1}[n]\rangle(p_{3})+Q'(p_{4})+\bar{Q}(p_{5})$ in Fig.~\ref{YYFeny}, where $Q$ and $Q'$ stand for the heavy $c$- or $b$-quark, respectively. The remaining ten Feynman diagrams can be conveniently obtained by exchanging the position of the incident photons. The total hard scattering amplitude is
\begin{equation}
{\cal M}=\sum_{i=1}^{20} {\cal M}_i \;,
\end{equation}
in which the amplitudes ${\cal M}_{2n-1}$ with $n=(1,\cdots,10)$ can be directly read from Fig.~\ref{YYFeny}, and ${\cal M}_{2n}$ can be obtained from ${\cal M}_{2n-1}$ by exchanging the momenta of the incident two photons. More explicitly, we have
\begin{widetext}
\begin{eqnarray}
{\cal M}_{1} &=& i{\cal C}\;\bar{u}_{s'}(p_4) \not\!\varepsilon^{\lambda_2}_{k_2} \frac{{\not\!p}_4-{\not\!k}_2+m_{Q'}}{(p_4-k_2)^2-m_{Q'}^2} \not\!\varepsilon^{\lambda_1}_{k_1} \frac{-{\not\!p}_3-{\not\!p}_5+m_{Q'}}{(p_3+p_5)^2-m_{Q'}^2} \gamma^{\sigma} \frac{\Pi(p_3)}{(p_5+p_{31})^2} \gamma_{\sigma} v_{s}(p_5) \;, \label{eq1} \\
{\cal M}_{2} &=& i{\cal C}\;\bar{u}_{s'}(p_4) \not\!\varepsilon^{\lambda_1}_{k_1} \frac{{\not\!p}_4-{\not\!k}_1+m_{Q'}}{(p_4-k_1)^2-m_{Q'}^2} \not\!\varepsilon^{\lambda_2}_{k_2} \frac{-{\not\!p}_3-{\not\!p}_5+m_{Q'}}{(p_3+p_5)^2-m_{Q'}^2} \gamma^{\sigma} \frac{\Pi(p_3)}{(p_5+p_{31})^2} \gamma_{\sigma} v_{s}(p_5) \;, \\
{\cal M}_{3} &=& i{\cal C}\;\bar{u}_{s'}(p_4) \not\!\varepsilon^{\lambda_2}_{k_2} \frac{{\not\!p}_4-{\not\!k}_2+m_{Q'}}{(p_4-k_2)^2-m_{Q'}^2} \gamma^{\sigma}  \frac{{\not\!k}_1-{\not\!p}_{32}+m_{Q'}}{(k_1-p_{32})^2-m_{Q'}^2} \not\!\varepsilon^{\lambda_1}_{k_1} \frac{\Pi(p_3)}{(p_5+p_{31})^2} \gamma_{\sigma} v_{s}(p_5) \;, \\
{\cal M}_4 &=& i{\cal C}\;\bar{u}_{s'}(p_4) \not\!\varepsilon^{\lambda_1}_{k_1} \frac{{\not\!p}_4-{\not\!k}_1+m_{Q'}}{(p_4-k_1)^2-m_{Q'}^2} \gamma^{\sigma}  \frac{{\not\!k}_2-{\not\!p}_{32}+m_{Q'}}{(k_2-p_{32})^2-m_{Q'}^2} \not\!\varepsilon^{\lambda_2}_{k_2} \frac{\Pi(p_3)}{(p_5+p_{31})^2} \gamma_{\sigma} v_{s}(p_5) \;,
\end{eqnarray}
\begin{eqnarray}
{\cal M}_5 &=& i{\cal C}\;\bar{u}_{s'}(p_4) \gamma^{\sigma} \frac{{\not\!p}_5+{\not\!p}_4+{\not\!p}_{31}+m_Q}{(p_5+p_4+p_{31})^2-m_Q^2} \not\!\varepsilon^{\lambda_2}_{k_2} \frac{{\not\!k}_1-{\not\!p}_{32}+m_{Q'}}{(k_1-p_{32})^2-m_{Q'}^2} \not\!\varepsilon^{\lambda_1}_{k_1} \frac{\Pi(p_3)}{(p_5+p_{31})^2} \gamma_{\sigma} v_{s}(p_5) \;,\\
{\cal M}_6 &=& i{\cal C}\;\bar{u}_{s'}(p_4) \gamma^{\sigma} \frac{{\not\!p}_5+{\not\!p}_4+{\not\!p}_{31}+m_Q}{(p_5+p_4+p_{31})^2-m_Q^2} \not\!\varepsilon^{\lambda_1}_{k_1} \frac{{\not\!k}_2-{\not\!p}_{32}+m_{Q'}}{(k_2-p_{32})^2-m_{Q'}^2} \not\!\varepsilon^{\lambda_2}_{k_2} \frac{\Pi(p_3)}{(p_5+p_{31})^2} \gamma_{\sigma} v_{s}(p_5) \;, \\
{\cal M}_{7} &=& i{\cal C}\;\bar{u}_{s'}(p_4) \not\!\varepsilon^{\lambda_2}_{k_2} \frac{{\not\!p}_4-{\not\!k}_2+m_{Q'}}{(p_4-k_2)^2-m_{Q'}^2} \gamma^{\sigma} \frac{\Pi(p_3)}{(k_1-p_5-p_{31})^2} \gamma_{\sigma} \frac{{\not\!k}_1-{\not\!p}_5+m_{Q}}{(k_1-p_5)^2-m_{Q}^2} \not\!\varepsilon^{\lambda_1}_{k_1} v_{s}(p_5)\;, \\
{\cal M}_{8} &=& i{\cal C}\;\bar{u}_{s'}(p_4) \not\!\varepsilon^{\lambda_1}_{k_1} \frac{{\not\!p}_4-{\not\!k}_1+m_{Q'}}{(p_4-k_1)^2-m_{Q'}^2} \gamma^{\sigma} \frac{\Pi(p_3)}{(k_2-p_5-p_{31})^2} \gamma_{\sigma} \frac{{\not\!k}_2-{\not\!p}_5+m_{Q}}{(k_2-p_5)^2-m_{Q}^2} \not\!\varepsilon^{\lambda_2}_{k_2} v_{s}(p_5)\;,
\end{eqnarray}
\begin{eqnarray}
{\cal M}_{9} &=& i{\cal C}\;\bar{u}_{s'}(p_4) \gamma^{\sigma} \frac{{\not\!k}_2-{\not\!p}_{32}+m_{Q'}}{(k_2-p_{32})^2-m_{Q'}^2} \not\!\varepsilon^{\lambda_2}_{k_2} \frac{\Pi(p_3)}{(k_2-p_4-p_{32})^2} \not\!\varepsilon^{\lambda_1}_{k_1} \frac{{\not\!p}_{31}-{\not\!k}_1+m_{Q}}{(p_{31}-k_1)^2-m_{Q}^2} \gamma_{\sigma} v_{s}(p_5) \;,\\
{\cal M}_{10} &=& i{\cal C}\;\bar{u}_{s'}(p_4) \gamma^{\sigma} \frac{{\not\!k}_1-{\not\!p}_{32}+m_{Q'}}{(k_1-p_{32})^2-m_{Q'}^2} \not\!\varepsilon^{\lambda_1}_{k_1} \frac{\Pi(p_3)}{(p_4-k_1+p_{32})^2} \not\!\varepsilon^{\lambda_2}_{k_2} \frac{{\not\!p}_{31}-{\not\!k}_2+m_{Q}}{(p_3-k_2)^2-m_{Q}^2} \gamma_{\sigma} v_{s}(p_5)\;, \\
{\cal M}_{11} &=& i{\cal C}\;\bar{u}_{s'}(p_4) \gamma^{\sigma} \frac{\Pi(p_3)}{(p_4+p_{32})^2} \not\!\varepsilon^{\lambda_2}_{k_2} \frac{{\not\!p}_{31}-{\not\!k}_2+m_Q}{(p_{31}-k_2)^2-m_Q^2} \not\!\varepsilon^{\lambda_1}_{k_1} \frac{-{\not\!p}_5-{\not\!p}_4-{\not\!p}_{32}+m_Q}{(p_5+p_4+p_{32})^2-m_Q^2} \gamma_{\sigma} v_{s}(p_5) \;, \\
{\cal M}_{12} &=& i{\cal C}\;\bar{u}_{s'}(p_4) \gamma^{\sigma} \frac{\Pi(p_3)}{(p_4+p_{32})^2} \not\!\varepsilon^{\lambda_1}_{k_1} \frac{{\not\!p}_{31}-{\not\!k}_1+m_Q}{(p_{31}-k_1)^2-m_Q^2} \not\!\varepsilon^{\lambda_2}_{k_2} \frac{-{\not\!p}_5-{\not\!p}_4-{\not\!p}_{32}+m_Q}{(p_5+p_4+p_{32})^2-m_Q^2} \gamma_{\sigma} v_{s}(p_5) \;,
\end{eqnarray}
\begin{eqnarray}
{\cal M}_{13} &=& i{\cal C}\;\bar{u}_{s'}(p_4) \gamma^{\sigma} \frac{\Pi(p_3)}{(p_4+p_{32})^2} \not\!\varepsilon^{\lambda_2}_{k_2} \frac{{\not\!p}_{31}-{\not\!k}_2+m_Q}{(p_{31}-k_2)^2-m_Q^2} \gamma_{\sigma} \frac{{\not\!k}_1-{\not\!p}_5+m_Q}{(k_1-p_5)^2-m_Q^2} \not\!\varepsilon^{\lambda_1}_{k_1} v_{s}(p_5) \;,\\
{\cal M}_{14} &=& i{\cal C}\;\bar{u}_{s'}(p_4) \gamma^{\sigma} \frac{\Pi(p_3)}{(p_4+p_{32})^2} \not\!\varepsilon^{\lambda_1}_{k_1} \frac{{\not\!p}_{31}-{\not\!k}_1+m_Q}{(p_{31}-k_1)^2-m_Q^2} \gamma_{\sigma} \frac{{\not\!k}_2-{\not\!p}_5+m_Q}{(k_2-p_5)^2-m_Q^2} \not\!\varepsilon^{\lambda_2}_{k_2} v_{s}(p_5) \;, \\
{\cal M}_{15} &=& i{\cal C}\;\bar{u}_{s'}(p_4) \gamma^{\sigma} \frac{\Pi(p_3)}{(p_4+p_{32})^2} \gamma_{\sigma} \frac{{\not\!p}_3+{\not\!p}_4+m_Q}{(p_3+p_4)^2-m_Q^2} \not\!\varepsilon^{\lambda_2}_{k_2} \frac{{\not\!k}_1-{\not\!p}_5+m_Q}{(k_1-p_5)^2-m_Q^2} \not\!\varepsilon^{\lambda_1}_{k_1} v_{s}(p_5) \;, \\
{\cal M}_{16} &=& i{\cal C}\;\bar{u}_{s'}(p_4) \gamma^{\sigma} \frac{\Pi(p_3)}{(p_4+p_{32})^2} \gamma_{\sigma} \frac{{\not\!p}_3+{\not\!p}_4+m_Q}{(p_3+p_4)^2-m_Q^2} \not\!\varepsilon^{\lambda_1}_{k_1} \frac{{\not\!k}_2-{\not\!p}_5+m_Q}{(k_2-p_5)^2-m_Q^2} \not\!\varepsilon^{\lambda_2}_{k_2} v_{s}(p_5) \;,
\end{eqnarray}
\begin{eqnarray}
{\cal M}_{17} &=& i{\cal C}\;\bar{u}_{s'}(p_4) \gamma^{\sigma} \frac{{\not\!k}_2-{\not\!p}_{32}+m_{Q'}}{(k_2-p_{32})^2-m_{Q'}^2} \not\!\varepsilon^{\lambda_2}_{k_2} \frac{\Pi(p_3)}{(k_1-p_5-p_{31})^2} \gamma_{\sigma} \frac{{\not\!k}_1-{\not\!p}_5+m_{Q}}{(k_1-p_5)^2-m_{Q}^2} \not\!\varepsilon^{\lambda_1}_{k_1} v_{s}(p_5) \;, \\
{\cal M}_{18} &=& i{\cal C}\;\bar{u}_{s'}(p_4) \not\!\varepsilon^{\lambda_1}_{k_1} \frac{{\not\!p}_4-{\not\!k}_1+m_{Q'}}{(p_4-k_1)^2-m_{Q'}^2} \gamma^{\sigma} \frac{\Pi(p_3)}{(p_4-k_1-p_{32})^2} \not\!\varepsilon^{\lambda_2}_{k_2} \frac{{\not\!p}_{31}-{\not\!k}_2+m_{Q}}{(p_{31}-k_2)^2-m_{Q}^2} \gamma_{\sigma} v_{s}(p_5)\;, \\
{\cal M}_{19} &=& i{\cal C}\;\bar{u}_{s'}(p_4) \not\!\varepsilon^{\lambda_2}_{k_2} \frac{{\not\!p}_4-{\not\!k}_2+m_{Q'}}{(p_4-k_2)^2-m_{Q'}^2} \gamma^{\sigma} \frac{\Pi(p_3)}{(p_4-k_2-p_{32})^2} \not\!\varepsilon^{\lambda_1}_{k_1} \frac{{\not\!p}_{31}-{\not\!k}_1+m_{Q}}{(p_{31}-k_1)^2-m_{Q}^2} \gamma_{\sigma} v_{s}(p_5)\;, \\
{\cal M}_{20} &=& i{\cal C}\;\bar{u}_{s'}(p_4) \gamma^{\sigma} \frac{{\not\!k}_1-{\not\!p}_{32}+m_{Q'}}{(k_1-p_{32})^2-m_{Q'}^2} \not\!\varepsilon^{\lambda_1}_{k_1} \frac{\Pi(p_3)}{(k_2-p_5-p_{31})^2} \gamma_{\sigma} \frac{{\not\!k}_2-{\not\!p}_5+m_{Q}}{(k_2-p_5)^2-m_{Q}^2} \not\!\varepsilon^{\lambda_2}_{k_2} v_{s}(p_5). \label{eq20}
\end{eqnarray}
\end{widetext}
Here, $\varepsilon^{\lambda_1}_{k_1}$ ($\varepsilon^{\lambda_2}_{k_2}$) is the polarization vector of the initial photon with momentum $k_1$ ($k_2$) and helicity state $\lambda_1$ ($\lambda_2$). $m_Q$ and $m_{Q'}$ are the masses of $Q$- and $Q'$-quark, respectively. ${\cal C}$ is the overall constant, ${\cal C}=\frac{4}{3}\delta^{ab}e^2 g^2 Q_e^2 $, where $Q_e$ equal $-\frac{1}{3}$ or $\frac{2}{3}$ for $b$- and $c$-quark, respectively, $a$ and $b$ are color indices for the final out-going quarks. $\Pi(p_3)$ stands for the spin-projection operator which depicts the $(Q\bar{Q'})$-pair evolving into the heavy quarkonium,
\begin{eqnarray}
\Pi(p_3) = \frac{1}{2\sqrt{M_{Q\bar{Q'}}}} [{\cal C}_1 \gamma^5 + {\cal C}_2 \not\!\varepsilon(s_z)] ({\not\!p}_{3}+M_{Q\bar{Q'}}) \;,
\end{eqnarray}
where ${\cal C}_1$=1 and ${\cal C}_2$=0 for the pseudoscalar state $|[Q\bar{Q'}]_{\bf 1}(^1S_0)\rangle$, ${\cal C}_1$=0 and ${\cal C}_2$=1 for the vector state $|[Q\bar{Q'}]_{\bf 1}(^3S_1)\rangle$. $\varepsilon(s_z)$ is the polarization vector for the vector state. The heavy quarkonium mass $M_{Q\bar{Q'}} = m_{Q}+m_{Q'}$ and the momenta of the constituent quarks in the bound system can be expressed as
\begin{displaymath}
p_{31} = \frac{m_Q}{M_{Q\bar{Q'}}}p_3 \;\; {\rm and}\;\;
p_{32} = \frac{m_{Q'}}{M_{Q\bar{Q'}}}p_3 \;.
\end{displaymath}

\subsection{The improved helicity amplitude approach}

In the present subsection, we adopt the improved helicity amplitude approach~\cite{helicity1} to deal with the hard scattering amplitude. The key point is to transform the Dirac spinor for the massive fermion (momentum $p$ with mass $m$) into the spinor of the massless fermions, i.e.
\begin{eqnarray}
u_{\pm\frac{1}{2}}(p) &=& \frac{1}{\sqrt{2p \cdot q}}({\not\!p}+m)|q_{\pm}\rangle \; , \label{spinor1} \\
v_{\pm\frac{1}{2}}(p) &=& \frac{1}{\sqrt{2p \cdot q}}({\not\!p}-m)|q_{\mp}\rangle \; ,  \label{spinor2}
\end{eqnarray}
where $|q_{\pm}\rangle$ is a massless fermion spinor with an arbitrary light-like momentum $q$ and helicity $\pm1$, which satisfies
\begin{eqnarray}\label{defml}
{\not\!q}|q\rangle = 0, \;\;\; |q_{\pm}\rangle = \omega_{\pm}|q\rangle \;,
\end{eqnarray}
where $\omega_{\pm}=\frac{1\pm \gamma^5}{2}$. The polarization vector of the photon $\varepsilon^{\pm}_{\mu}$ with momentum $k$ has the form related to the reference light-like momentum $q$ as follows,
\begin{eqnarray}
\varepsilon^{\pm}_{\mu}(k,q) &=& \pm\frac{\langle k_{\pm}|\gamma_{\mu}|q_{\pm}\rangle}{\sqrt{2} \langle q_{\mp}|k_{\pm} \rangle} \; ,\\
\not\!\varepsilon^{\pm}(k,q) &=& \pm\frac{\sqrt{2}}{\langle q_{\mp}|k_{\pm} \rangle}(|k_{\mp}\rangle\langle q_{\mp}|+|q_{\pm}\rangle\langle k_{\pm}|) \; ,
\end{eqnarray}
where, the $\langle q_{\mp}|k_{\pm} \rangle$ denotes the spinor inner-product.

The amplitude ${\cal M}_i$ with $i=(1,\cdots,20)$ can be factorized into two parts. One part is the process with free final quarks (all of which are on shell), i.e. $\gamma \gamma \to \bar{Q'} + Q + \bar{Q} + Q'$, and the other part is the free $Q$ and $\bar{Q'}$ binding into the required Fock state, $Q+\bar{Q'} \to |(Q\bar{Q'})[n]\rangle$. With the help of the introduction of massless fermion spinors as defined in Eq.~(\ref{spinor1}) and (\ref{spinor2}), the amplitude with explicit helicity states for all the particles' helicities can be formulated as
\begin{widetext}
\begin{eqnarray}\label{freeM}
{\cal M}_i^{(\lambda_1,\lambda_2,\lambda_5,\lambda_6)} (k_1,k_2,p_{31},p_{32},p_4,p_5) = {\cal C} \sum\limits_{\lambda_3,\lambda_4} D_1 M_{Fi}^{(\lambda_1,\lambda_2,\lambda_3,\lambda_4, \lambda_5,\lambda_6)}(k_1,k_2,p_{31},p_{32},p_4,p_5)\times M_{BS}^{(\lambda_3,\lambda_4)}(p_{31},p_{32})\; ,
\end{eqnarray}
\end{widetext}
where $D_1$ is the normalization factor from the transformation between massive and massless fermion spinor, which is defined as
\begin{eqnarray}
D_1 &=& \frac{1}{\sqrt{2 p_{31} \cdot q}} \frac{1}{\sqrt{2 p_{32} \cdot q}} \frac{1}{\sqrt{2 p_{4} \cdot q}} \frac{1}{\sqrt{2 p_{5} \cdot q}} \;.
\end{eqnarray}
The amplitudes $M_{BS}^{(\lambda_3,\lambda_4)}(p_{31},p_{32})$ for the bound state part are simpler and can be expressed as
\begin{equation}\label{BS}
M_{BS}^{(\lambda_3,\lambda_4)}(p_{31},p_{32}) = D_2 \langle q_{\lambda_4}|(a \gamma^5 + b\not\!\varepsilon(s_z)) \frac{{\not\!p}_{3}+M_{Q\bar{Q'}}} {2\sqrt{M_{Q\bar{Q'}}}}|q_{\lambda_3}\rangle \;.
\end{equation}
where $D_2$ is the normalization factor from the binding system, and we have
\begin{eqnarray}
D_2 &=& \frac{1}{\sqrt{2 p_{31} \cdot q}} \frac{1}{\sqrt{2 p_{32} \cdot q}} \;.
\end{eqnarray}
With the help of the relation ${\not\!p}=|p_+\rangle\langle p_+|+|p_-\rangle\langle p_-|$, the amplitudes $M_{BS}^{(\lambda_3,\lambda_4)}(p_{31},p_{32})$ can be easily simplified as
\begin{eqnarray}
M_{(^1S_0)}^{(\lambda_3,\lambda_4)}(p_{31},p_{32}) = \frac{\sqrt{M_{Q\bar{Q'}}}}{2\sqrt{m_Q m_{Q'}}} \delta_{\lambda_3 \lambda_4}(\delta_{\lambda_4-}-\delta_{\lambda_4+}),
\end{eqnarray}
and
\begin{widetext}
\begin{eqnarray}
M_{(^3S_1)}^{(\lambda_3,\lambda_4)}(p_{31},p_{32}) = \frac{\sqrt{M_{Q\bar{Q'}}}}{2\sqrt{m_Q m_{Q'}}}
 \Bigg[ \delta_{\lambda_3 \lambda_4}(\delta_{\lambda_4+}+\delta_{\lambda_4-})  \left( \frac{M_{Q\bar{Q'}} \varepsilon(s_z) \cdot q}{p_3 \cdot q} \right) + \left(\frac{1}{2p_3 \cdot q}\right)\langle q_{\lambda_4}|\not\!\varepsilon(s_z){\not\!p}_{3}|q_{\lambda_{3}}\rangle \Bigg] ,
\end{eqnarray}
\end{widetext}
for the $^1S_0$ and $^3S_1$ states, respectively, $\delta$ being the usual Kronecker symbol.

The amplitude $M_{Fi}^{(\lambda_1,\lambda_2, \lambda_3,\lambda_4, \lambda_5,\lambda_6)}(k_1,k_2,p_{31},p_{32},p_4,p_5)$ for the free quark part, $\gamma(k_1,\lambda_1)+\gamma(k_2,\lambda_2) \to Q(p_{31},\lambda_3) + \bar{Q'}(p_{32},\lambda_4) + Q'(p_4,\lambda_5) + \bar{Q}(p_5,\lambda_6)$, can be written as
\begin{widetext}
\begin{eqnarray}
&&M_{Fi}^{(\lambda_1,\lambda_2,\lambda_3,\lambda_4,\lambda_5,\lambda_6)} (k_1,k_2,p_{31},p_{32},p_4,p_5) \;\nonumber\\
&=& X_i \times \langle q_{\lambda_5}|({\not\!p}_{4}+m_{Q'}) \cdot \Gamma_{1i} \cdot ({\not\!p}_{32}-m_{Q'})|q_{\lambda_4}\rangle \times \langle q_{\lambda_3}|({\not\!p}_{31}+m_{Q}) \cdot \Gamma_{2i} \cdot ({\not\!p}_5-m_{Q})|q_{\lambda_6}\rangle \;.
\end{eqnarray}
\end{widetext}
$\Gamma_{1i,2i}$ are Dirac $\gamma$-matrix strings related to the $i_{\rm th}$-diagram, which include the momentums $k_1$, $k_2$ and the helicities $\lambda_1$, $\lambda_2$ of the initial photons. $X_i$ is the scalar product terms from all the propagators of the $i_{\rm th}$ diagram. Both $\Gamma_{1i,2i}$ and $X_i$ can be read from ${\cal M}_{i}$ as listed in Eqs.~(\ref{eq1},$\cdots$,\ref{eq20}). Every amplitude $M_{Fi}^{(\lambda_1,\lambda_2,\lambda_3,\lambda_4,\lambda_5,\lambda_6)} (k_1,k_2,p_{31},p_{32},p_4,p_5)$ is constructed by two fermion lines. It is found that those twenty amplitudes can be constructed by six ``basic functions" denoted by $E_{m,j,k}(k_1,k_2,p_{31},p_{32},p_4,p_5)$ $(m=1,2,\cdots,6; j=1,\cdots,4)$. The subscript $k$ equals $2^6=64$ possible helicity combinations of $(\lambda_1,\lambda_2,\lambda_3, \lambda_4, \lambda_5, \lambda_6)$. Here $j=(1,\cdots,4)$ stands for a specific type of interchange: $j=1$ means identical (without any interchange), $j=2$ means interchange of the two photons, $j=3$ means interchange of the quark ($Q$) and the anti-quark ($\bar{Q'}$), and $j=4$ means interchange of the photons and the quark and anti-quark. The six basic functions for $j=1$ can be expressed as follows
\begin{widetext}
\begin{eqnarray}
E_{1,1,k} &=& X_{16,1} \cdot f_1(p_{32},p_4,\lambda_4,\lambda_5)\cdot f_4(k_1,k_2,p_{31},p_5,\lambda_1,\lambda_2,\lambda_3,\lambda_6) \;, \nonumber\\
E_{2,1,k} &=& X_{14,1} \cdot f_1(p_{32},p_4,\lambda_4,\lambda_5)\cdot f_5(k_1,k_2,p_{31},p_5,\lambda_1,\lambda_2,\lambda_3,\lambda_6)\;, \nonumber\\
E_{3,1,k} &=& X_{12,1} \cdot f_1(p_{32},p_4,\lambda_4,\lambda_5)\cdot f_6(k_1,k_2,p_{31},p_5,\lambda_1,\lambda_2,\lambda_3,\lambda_6)\;, \nonumber\\
E_{4,1,k} &=& X_{7,1} \cdot f_2(k_1,p_{31},p_5,\lambda_1,\lambda_3,\lambda_6)\cdot f_3(k_2,p_{32},p_4,\lambda_2,\lambda_4,\lambda_5)\;, \nonumber\\
E_{5,1,k} &=& X_{19,1} \cdot f_3(k_1,p_{31},p_5,\lambda_1,\lambda_3,\lambda_6)\cdot f_3(k_2,p_{32},p_4,\lambda_2,\lambda_4,\lambda_5)\;, \nonumber\\
E_{6,1,k} &=& X_{17,1} \cdot f_2(k_1,p_{31},p_5,\lambda_1,\lambda_3,\lambda_6)\cdot f_2(k_2,p_{32},p_4,\lambda_2,\lambda_4,\lambda_5)\;,
\end{eqnarray}
\end{widetext}
where $X_{i,j}$ stands for the transformation of $X_i$ by doing the $j_{\rm th}$-type of interchanges mentioned above. $f_{(1,2,\cdots,6)}$ are basic fermion lines corresponding to different types of Dirac-$\gamma$ structures
\begin{widetext}
\begin{eqnarray}
f_1(q_1,q_2,\lambda'_1,\lambda'_2) &=& \langle q_{\lambda'_1}|({\not\!q}_1+m)\gamma_{\rho}({\not\!q}_2-m)|q_{\lambda'_2}\rangle \;, \\
f_2(k,q_1,q_2,\lambda'_3,\lambda'_1,\lambda'_2) &=& \langle q_{\lambda'_1}|({\not\!q}_1+m)\gamma_{\rho}({\not\!k}-{\not\!q}_2+m)\not\!\varepsilon^{\lambda'_3}(k,q)({\not\!q}_2-m)|q_{\lambda'_2}\rangle \;, \\
f_3(k,q_1,q_2,\lambda'_3,\lambda'_1,\lambda'_2) &=& \langle q_{\lambda'_1}|({\not\!q}_1+m)\not\!\varepsilon^{\lambda'_3}(k,q)({\not\!q}_1-{\not\!k}+m)\gamma_{\rho}({\not\!q}_2-m)|q_{\lambda'_2}\rangle \;,  \\
f_4(k,k',q_1,q_2,\lambda'_3,\lambda'_4,\lambda'_1,\lambda'_2) &=& \langle q_{\lambda'_1}|({\not\!q}_1+m)\gamma_{\rho}({\not\!k}+{\not\!k}'-{\not\!q}_2+m)\not\!\varepsilon^{\lambda'_3}(k,q)({\not\!k}'-{\not\!q}_2+m)
\not\!\varepsilon^{\lambda'_4}(k',q)({\not\!q}_2-m)|q_{\lambda'_2}\rangle \;, \\
f_5(k,k',q_1,q_2,\lambda'_3,\lambda'_4,\lambda'_1,\lambda'_2) &=& \langle q_{\lambda'_1}|({\not\!q}_1+m)\not\!\varepsilon^{\lambda'_3}(k,q)({\not\!q}_1-{\not\!k}+m)\gamma_{\rho}({\not\!k}'-{\not\!q}_2+m)
\not\!\varepsilon^{\lambda'_4}(k',q)({\not\!q}_2-m)|q_{\lambda'_2}\rangle \;, \\
f_6(k,k',q_1,q_2,\lambda'_3,\lambda'_4,\lambda'_1,\lambda'_2) &=& \langle q_{\lambda'_1}|({\not\!q}_1+m)\not\!\varepsilon^{\lambda'_3}(k,q)({\not\!q}_1-{\not\!k}+m)\not\!\varepsilon^{\lambda'_4}(k',q)({\not\!q}_1-{\not\!k}-{\not\!k}'+m)
\gamma_{\rho}({\not\!q}_2-m)|q_{\lambda'_2}\rangle \;,
\end{eqnarray}
where $q_1^2=q_2^2=m^2$ and $q^2=k^2=k'^2=0$. The amplitude for $\gamma(k_{1})\gamma(k_{2}) \to |(Q\bar{Q'})_{\bf 1}[n]\rangle(p_{3})+Q'(p_{4})+\bar{Q}(p_{5})$ can be written as
\begin{eqnarray}
&&{\cal M}^{(\lambda_1,\lambda_2,\lambda_5,\lambda_6)}(k_1,k_2,p_{31},p_{32},p_4,p_5) = \sum\limits_{i=1}^{20} {\cal M}_i^{(\lambda_1,\lambda_2,\lambda_5,\lambda_6)}(k_1,k_2,p_{31},p_{32},p_4,p_5) \;\nonumber\\
&=& {\cal C} D_1 \sum\limits_{\lambda_3,\lambda_4} M_{BS}^{(\lambda_3,\lambda_4)}(p_{31},p_{32})\left[\sum\limits_{m=1}^{6} \sum\limits_{j=1}^{2}E_{m,j,k}(k_1,k_2,p_{31},p_{32},p_4,p_5)+\sum\limits_{m=1}^{4} \sum\limits_{j=3}^{4}E_{m,j,k}(k_1,k_2,p_{31},p_{32},p_4,p_5) \right],
\end{eqnarray}
\end{widetext}
where the subscript $k$ represents the indices $(\lambda_1, \lambda_2, \lambda_3, \lambda_4, \lambda_5, \lambda_6)$, of which $\lambda_3$ and $\lambda_4$ should be summed over. All the functions $E_{m,j,k}$, with $m$ and $j$ fixed, are related to each other by proper complex conjugation with or without changing the overall sign. As a final step, what we need is to numerically calculate those six basic fermion structures $f_i$ under specific helicities.

We take $f_1(q_1,q_2,+,+)$ as an explicit example to explain how to do the simplification. By introducing another light-like momentum $q'_i$ expressed in terms of $q_i$ and the reference momentum $q$ as
\begin{eqnarray}
q'_i &=& q_i - \frac{q_i^2}{2q_i \cdot q} q\;(i=1,2)\;,
\end{eqnarray}
we obtain
\begin{eqnarray}
f_1(q_1,q_2,+,+) &=& \langle q_{+}|q'_{1-}\rangle\langle q'_{1-}|\gamma_{\rho}|q'_{2-}\rangle\langle q'_{2-}|q_{+}\rangle \nonumber\\
&& -m^2 \langle q_{+}|\gamma_{\rho}|q_{+}\rangle \;.
\end{eqnarray}
Using the definitions (\ref{defml}) together with the formulae
\begin{eqnarray}
\langle p_+|{\not\!k}_1 \cdots {\not\!k}_n|q_+\rangle = \langle q_-|{\not\!k}_n \cdots {\not\!k}_1|p_-\rangle \;,\nonumber
\end{eqnarray}
where $n$ is an odd integer, we finally obtain
\begin{eqnarray}
f_1(q_1,q_2,+,+) &=& \langle q'_{1+}|{\not\!q}|q'_{2+}\rangle\langle q'_{2+}|\gamma_{\rho}|q'_{1+}\rangle \nonumber\\
&& - m^2 \langle q_{+}|\gamma_{\rho}|q_{+}\rangle \;.
\end{eqnarray}
Following similar procedures, one can simplify all basic fermion lines $f_{(1,2,\cdots,6)}$. There are $2^2=4$ helicity combinations for $f_1$, $2^3=8$ for $f_{2,3}$, and $2^4=16$ for $f_{4,5,6}$. We note that the basic fermion lines are finally transformed into fundamental elements, i.e. the spinor products $\langle q'_{1+}|{\not\!q}|q'_{2+}\rangle$ and inner products $\langle p_{\mp}|q_{\pm}\rangle$. Terms like $\langle q'_{2+}|\gamma_{\rho}|q'_{1+}\rangle$ can be expressed in terms of inner products after Lorentz-index contracting with the help of the Fierz rearrangement theorem. The basic spinor products and inner products can be done numerically~\cite{helicity1}. For self-consistency, we put the evaluations of those basic elements in the Appendix.

As a cross check of the improved helicity amplitude approach, we also adopt the improved trace technology~\cite{itt0,zbc0,tbc1,zbc1} to deal with the hard scattering amplitude at the amplitude level. Under the improved trace technology, the hard-scattering amplitude can be directly written as a trace form and be expressed by dot products of the known particle momenta as that of the squared amplitude. Thus, we can also get the numerical results for the hard scattering amplitudes at the amplitude level. Numerically, we find that the results for the cross sections from those two approaches are the same under the same input parameters.

\section{Numerical results}\label{results}

\subsection{Input parameters}

The $b$-quark mass is taken as $m_b=4.9$ GeV and the $c$-quark mass as $m_c=1.5$ GeV. The quarkonium mass $M_{Q\bar{Q'}}$ is chosen as the sum of the constituent quark masses so as to ensure the gauge invariance of the hard-scattering amplitude, e.g., $M_{c\bar{c}}=2m_c$, $M_{b\bar{b}}=2m_b$, and $M_{c\bar{b}}=m_b+m_c$. The fine-structure constant is fixed as $\alpha=1/137$. We set the renormalization and factorization scales to be the transverse mass of the final bound state, i.e., $\mu_r=\mu_f=M_t=\sqrt{M_{Q\bar{Q'}}^2+p_t^2}$. The coupling constant is running at the leading order. The $\Lambda_{\rm QCD}$ is fixed by the measured value of $\alpha_s(m_Z)=0.1184$ with $m_Z=91.1876$ GeV~\cite{pdg2012}. As for the wavefunction at the origin $|\Psi^{Q\bar{Q'}}_{S}(0)| ={|R^{Q\bar{Q'}}_{S}(0)|}/{\sqrt{4\pi}}$, we adopt~\cite{BTpM}: $|R_S^{c\bar{c}}(0)|^2=0.810\;{\rm GeV}^3$, $|R_S^{c\bar{b}}(0)|^2=1.642\;{\rm GeV}^3$, and $|R_S^{b\bar{b}}(0)|^2=6.477\;{\rm GeV}^3$ for the $S$-wave $|c\bar{c}\rangle$, $|c\bar{b}\rangle$, and $|b\bar{b}\rangle$ bound states, respectively.

\subsection{Basic results}

\begin{table}[htb]
\begin{tabular}{|c|c|c|c|}
\hline
 ~~$\sqrt{S}$~~ & ~~250 GeV~~ & ~~500 GeV~~ & ~~1 TeV~~ \\
\hline
$\sigma_{\eta_{c}}({\rm fb}) $ & 668 & 278 & 107 \\
\hline
$\sigma_{\emph{J}/\psi}({\rm fb}) $ & 1229 & 537 & 215 \\
\hline
$\sigma_{B_c}({\rm fb}) $ & 15.6 & 8.27 & 3.80 \\
\hline
$\sigma_{B_c^*}({\rm fb}) $ & 90.3 & 43.0 & 18.3 \\
\hline
$\sigma_{\eta_b}({\rm fb}) $ & 1.72 & 0.90 & 0.40 \\
\hline
$\sigma_{\Upsilon}({\rm fb}) $ & 2.92 & 1.60 & 0.75 \\
\hline
\end{tabular}
\caption{Total cross sections for the heavy quarkonium photoproduction at the ILC. Three $e^+e^-$ collision energies, $\sqrt{S}=250{\rm GeV}, 500{\rm GeV}, 1{\rm TeV}$, are adopted. }\label{tcrs}
\end{table}

Total cross sections for the heavy quarkonium photoproduction at the ILC are presented in Table~\ref{tcrs}, where three collision energies, $\sqrt{S}=250{\rm GeV}$, $500{\rm GeV}$ and $1{\rm TeV}$ are adopted. It is noted that total cross sections decrease with the increment of $\sqrt{S}$, e.g.,
\begin{eqnarray}
&& \sigma_{|c\bar{c}\rangle}|_{250{\rm GeV}} : \sigma_{|c\bar{c}\rangle}|_{500{\rm GeV}} : \sigma_{|c\bar{c}\rangle}|_{1{\rm TeV}} \simeq 6:2:1 , \nonumber\\
&& \sigma_{|c\bar{b}\rangle}|_{250{\rm GeV}} : \sigma_{|c\bar{b}\rangle}|_{500{\rm GeV}} : \sigma_{|c\bar{b}\rangle}|_{1{\rm TeV}} \simeq 6:2:1 , \nonumber\\
&& \sigma_{|b\bar{b}\rangle}|_{250{\rm GeV}} : \sigma_{|b\bar{b}\rangle}|_{500{\rm GeV}} : \sigma_{|b\bar{b}\rangle}|_{1{\rm TeV}} \simeq 4:2:1  , \nonumber
\end{eqnarray}
where two $S$-wave states $^1S_0$ and $^3S_1$ have been summed up for the heavy quarkonium photoproduction. In the following, we adopt $\sqrt{S}=500{\rm GeV}$ to do our discussion.

At $\sqrt{S}=500$ GeV, when summing up both the $^{1}S_{0}$ and $^3S_1$ states' contributions together, we have $\sigma_{|c\bar{c}\rangle}=815$fb, $\sigma_{|c\bar{b}\rangle}=51.27$fb, and $\sigma_{|b\bar{b}\rangle}=2.5$fb for the $|c\bar{c}\rangle$, $|c\bar{b}\rangle$, and $|b\bar{b}\rangle$ bound states, respectively. Thus, we obtain $\sigma_{|c\bar{c}\rangle}:\sigma_{|c\bar{b}\rangle}:\sigma_{|b\bar{b}\rangle} =493:21:1$. The charmonium photoproduction cross section is larger than that of $(c\bar{b})$-quarkonium as well as the bottomonium by about two orders of magnitude. If setting the integrated luminosity as $10^{4}$fb$^{-1}$, we shall have $2.8\times 10^{6}$ $\eta_c$, $5.4\times 10^{6}$ $J/\psi$, $8.3\times 10^{4}$ $B_c$, $4.3\times 10^{5}$ $B^*_c$, $9.0\times 10^{3}$ $\eta_b$, and $1.6\times 10^{4}$ $\Upsilon$ events via the photoproduction channels. Thus, the photoproduction at ILC shall also be helpful for studying the properties of the heavy quarkonium.

\begin{figure}[htb]
\includegraphics[width=0.48\textwidth]{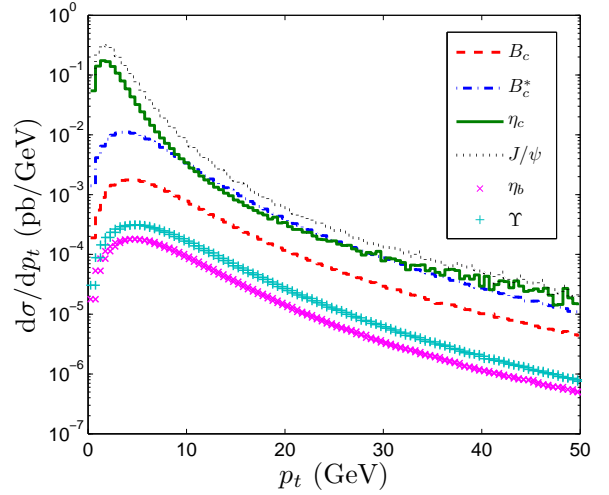}
\caption{Differential cross sections versus the transverse momentum ($p_t$) of the heavy quarkonium photoproduction at the ILC with $\sqrt{S}=500$ GeV. } \label{pt}
\end{figure}

\begin{figure}[htb]
\includegraphics[width=0.48\textwidth]{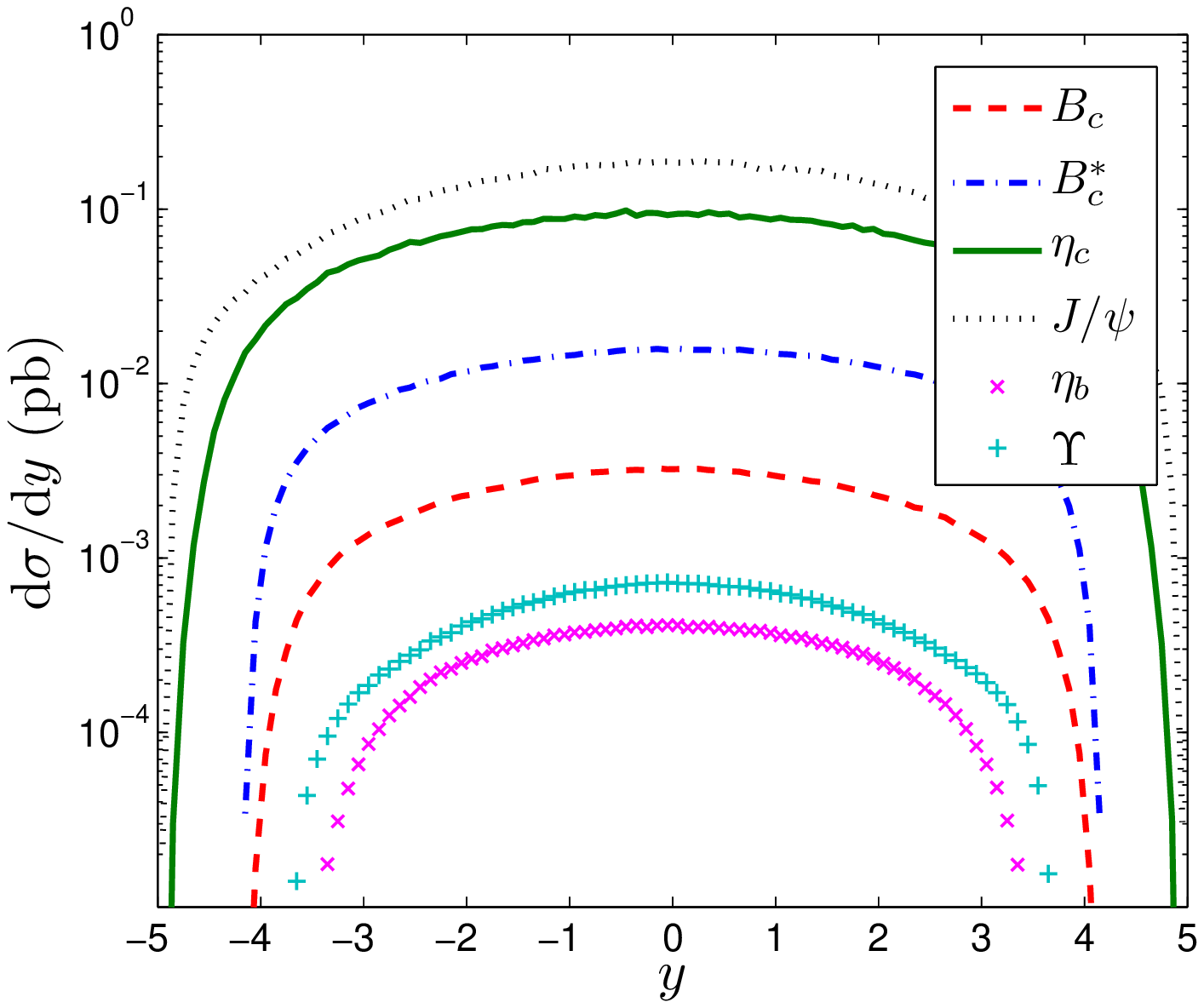}
\caption{The rapidity distributions of the heavy quarkonium photoproduction at the ILC with $\sqrt{S}=500$ GeV. } \label{rap}
\end{figure}

\begin{figure}[htb]
\includegraphics[width=0.48\textwidth]{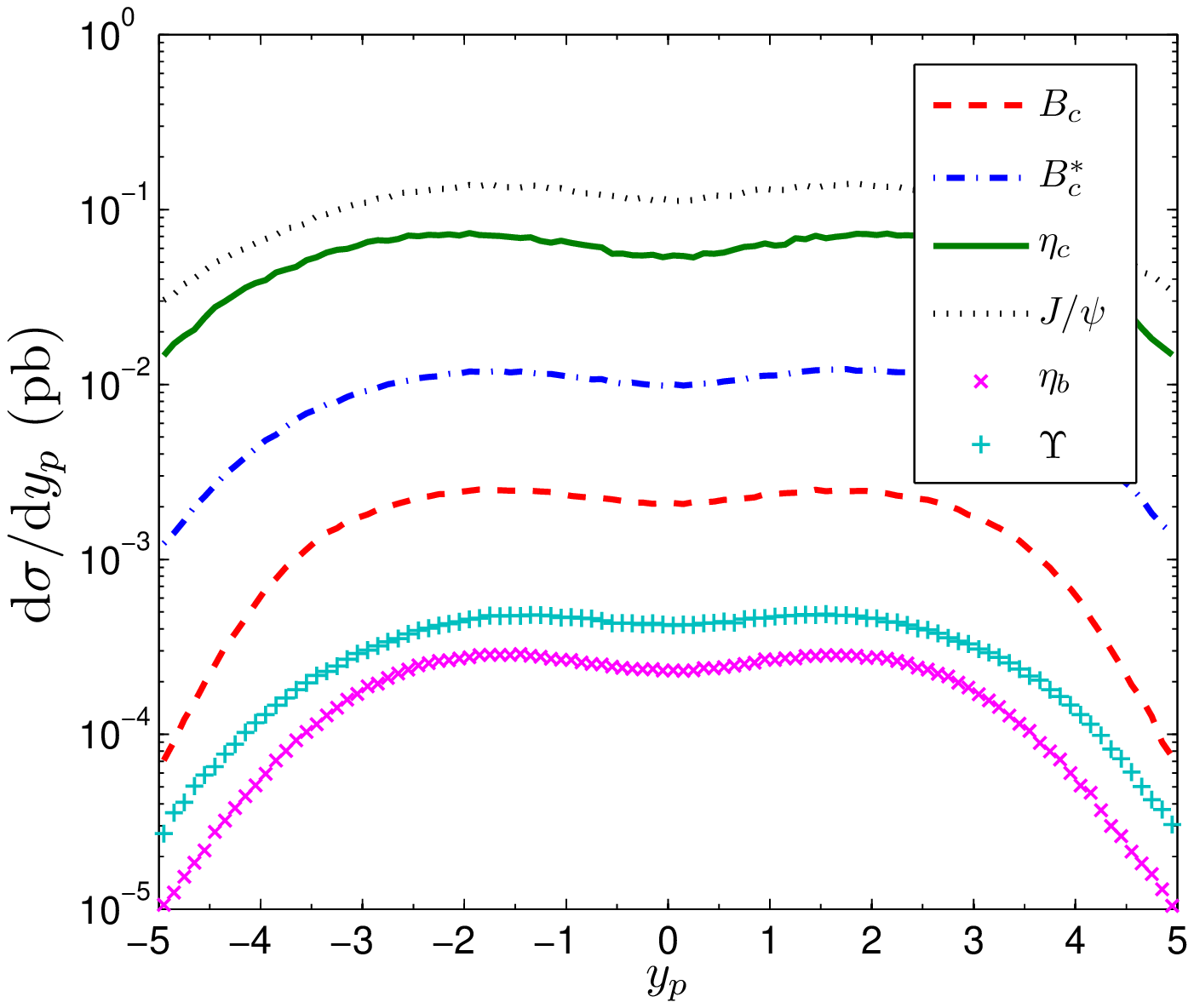}
\caption{The pseudorapidity distributions of the heavy quarkonium photoproduction at the ILC with $\sqrt{S}=500$ GeV.}\label{psrap}
\end{figure}

\begin{figure}[htb]
\includegraphics[width=0.48\textwidth]{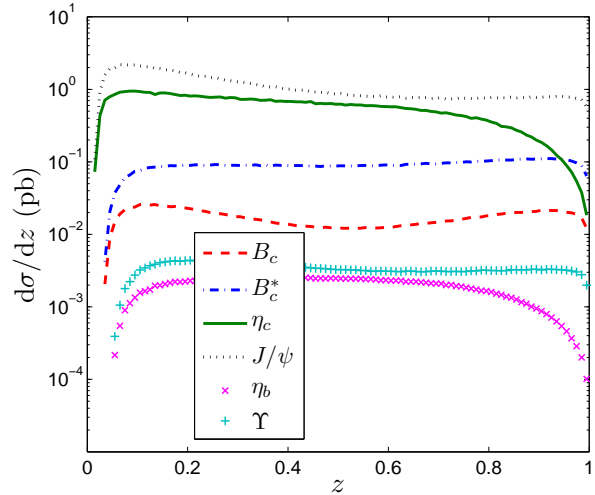}
\caption{Differential cross sections $d\sigma/dz$ versus $z$ for the heavy quarkonium photoproduction at the ILC with $\sqrt{S}=500$ GeV. } \label{yyz}
\end{figure}

We present the $p_t$ distributions for the heavy quarkonium photoproduction in Fig.~\ref{pt}. The $p_t$ distributions have a peak for $p_t$ around several GeV and drop down logarithmically in the large $p_t$ region. We draw rapidity ($y$) and pseudorapidity ($y_p$) distributions in Fig.~\ref{rap} and \ref{psrap}. There is a plateau within $|y|\lesssim 4$ for the charmonium photoproduction, $|y|\lesssim 3.5$ for the $B_c$ mesons photoproduction, and $|y|\lesssim 3$ for the bottomonium photoproduction.

\begin{table}[htb]
\begin{tabular}{|c|c|c|c|c|c|c|}
\hline
 & ~$\sigma_{\eta_{c}}$~ & ~$\sigma_{\emph{J}/\psi}$~ & ~$\sigma_{B_c} $~ & ~$\sigma_{B_c^*} $~ & ~$\sigma_{\eta_b} $~ & ~$\sigma_{\Upsilon} $~ \\
\hline
$p_t>1$ GeV& 227 & 451 & 8.09 & 41.7 & 0.88 & 1.57 \\
\hline
$p_t>2$ GeV& 145 & 299 & 7.58 & 38.0 & 0.83 & 1.48 \\
\hline
$p_t>3$ GeV& 84.9 & 181 & 6.79 & 32.8 & 0.76 & 1.35 \\
\hline
\end{tabular}
\caption{Total cross sections (in unit: fb) for the heavy quarkonium photoproduction at the ILC with $\sqrt{S}=500$GeV under various $p_t$ cuts. } \label{ptcuts}
\end{table}

In a high energy collider, the heavy quarkonium events with a small $p_t$ and/or a large rapidity $y$ cannot be measured directly. Therefore, events with proper kinematic cuts on $p_t$ and $y$ should be put in the estimates. Numerical results under several $p_t$ cuts are put in Table~\ref{ptcuts} and the results under several $y$ cuts are put in Table~\ref{ycuts}.

\begin{table}[htb]
\begin{tabular}{|c|c|c|c|c|c|c|}
\hline
 & ~$\sigma_{\eta_{c}} $~ & ~$\sigma_{\emph{J}/\psi} $~ & ~$\sigma_{B_c} $~ & ~$\sigma_{B_c^*} $~ & ~$\sigma_{\eta_b} $~ & ~$\sigma_{\Upsilon} $~ \\
\hline
\hline
$|y|<1$& 88.5 & 174 & 3.08 & 14.8 & 0.38 & 0.68 \\
\hline
$|y|<2$& 171 & 331 & 5.77 & 28.3 & 0.71 & 1.22 \\
\hline
$|y|<3$& 235 & 451 & 7.61 & 38.5 & 0.88 & 1.53 \\
\hline
\end{tabular}
\caption{Total cross sections (in unit: fb) for the photoproduction of heavy quarkonium with $\sqrt{S}=500$GeV under various rapidity cuts. } \label{ycuts}
\end{table}

As a final remark, we present the differential cross sections $d\sigma/dz$ versus $z$ in Fig.~\ref{yyz}, where $z =\frac{2}{\hat{s}}(k_1+k_2)\cdot p_3$ with $\hat{s}=x_1 x_2 S$ being the invariant mass of the initial photons of the subprocess. In the subprocess center-of-mass frame, $z$ is simply twice the fraction of the total energy carried by the heavy quarkonium and is experimentally observable.

\subsection{A discussion of theoretical uncertainties}

When discussing the uncertainty from one parameter, the other parameters shall be fixed to be their central values.

\begin{table}[htb]
\begin{tabular}{|c|c|c|c|}
\hline
 $m_{c}$ (GeV) & ~~$1.4$~~ & ~~$1.5$~~ & ~~$1.6$~~ \\
\hline
$\sigma_{\eta_{c}} $(fb) & 376 & 278 & 213 \\
\hline
$\sigma_{\emph{J}/\psi}$(fb) & 726 & 537 & 407 \\
\hline
$\sigma_{B_c}$(fb) & 9.82 & 8.27 & 7.07 \\
\hline
$\sigma_{B_c^*}$(fb) & 50.3 & 43.0 & 37.2 \\
\hline
\end{tabular}
\caption{Variations for the total cross-sections by taking $m_c=1.5\pm0.1$ GeV with $\sqrt{S}=500$ GeV. $m_b=4.9$GeV and $\mu_r=M_t$. }
\label{unmc}
\end{table}

\begin{table}[htb]
\begin{tabular}{|c|c|c|c|}
\hline
$m_{b}$ (GeV) & ~~$4.7$~~ & ~~$4.9$~~ & ~~$5.1$~~ \\
\hline
$\sigma_{B_c}$(fb) & 8.95 & 8.27 & 7.66 \\
\hline
$\sigma_{B_c^*}$(fb) & 46.8 & 43.0 & 39.7 \\
\hline
$\sigma_{\eta_b}$(fb) & 1.07 & 0.90 & 0.76 \\
\hline
$\sigma_{\Upsilon}$(fb) & 1.92 & 1.60 & 1.34 \\
\hline
\end{tabular}
\caption{Variations for the total cross-sections by taking $m_b=4.9\pm0.2$ GeV with $\sqrt{S}=500$ GeV. $m_c=1.5$GeV and $\mu_r=M_t$. }
\label{unmb}
\end{table}

To estimate the theoretical uncertainties for the heavy quarkonium photoproduction from the heavy quark masses, we take $m_c=1.50\pm0.10$ GeV and $m_b=4.9\pm0.20$ GeV. As shown in Table \ref{unmc}, at the ILC with $\sqrt{S}=500$ GeV, the uncertainties for $m_c=1.50\pm0.10$ GeV are
\begin{eqnarray}\label{cmass}
\sigma_{\eta_{c}} &=& 278^{+98}_{-65}\;{\rm fb}, \\
\sigma_{\emph{J}/\psi} &=& 537^{+189}_{-130}\;{\rm fb}, \\
\sigma_{B_c} &=& 8.27^{+1.55}_{-1.20}\;{\rm fb}, \\
\sigma_{B_c^*} &=& 43.0^{+7.30}_{-5.80}\;{\rm fb}.
\end{eqnarray}
Similarly, as shown in Table \ref{unmb}, the uncertainties for $m_b=4.9\pm0.20$ GeV are
\begin{eqnarray}\label{bmass}
\sigma_{B_c} &=& 8.27 ^{+0.68}_{-0.61} \;{\rm fb}, \\
\sigma_{B_c^*} &=& 43.0^{+3.8}_{-3.3}\;{\rm fb}, \\
\sigma_{\eta_b} &=& 0.90^{+0.17}_{-0.14} \;{\rm fb}, \\
\sigma_{\Upsilon} &=& 1.60^{+0.32}_{-0.26}\;{\rm fb}.
\end{eqnarray}
Tables \ref{unmc} and \ref{unmb} show that total cross sections decrease with increment of the $c$-quark or $b$-quark mass.

\begin{table}
\begin{tabular}{|c|c|c|c|c|c|c|}
\hline
  & ~$\sigma_{\eta_{c}}$~ & ~$\sigma_{\emph{J}/\psi}$~ & ~$\sigma_{B_c}$~ & ~$\sigma_{B_c^*}$~ & ~$\sigma_{\eta_b}$~ & ~$\sigma_{\Upsilon}$~ \\
\hline
$\mu_r=\sqrt{\hat{s}}$ & 108 & 201 & 3.92 & 21.1 & 0.47 & 0.83 \\
\hline
$\mu_r=\sqrt{\hat{s}}/2$ & 140 & 256 & 4.86 & 26.5 & 0.59 & 1.03 \\
\hline
$\mu_r=M_t$ & 278 & 537 & 8.27 & 43.0 & 0.90 & 1.60 \\
\hline
\end{tabular}
\caption{Total cross sections (in unit: fb) for the heavy quarkonium photoproduction under the conventional renormalization scale setting for three scale choices $\mu_r=\sqrt{\hat{s}}$, $\sqrt{\hat{s}}/2$, and $M_t$. $\sqrt{S}=500$ GeV.} \label{convQ}
\end{table}

\begin{table}
\begin{tabular}{|c|c|c|c|c|c|c|}
\hline
  & ~$\sigma_{\eta_{c}}$~ & ~$\sigma_{\emph{J}/\psi}$~ & ~$\sigma_{B_c}$~ & ~$\sigma_{B_c^*}$~ & ~$\sigma_{\eta_b}$~ & ~$\sigma_{\Upsilon}$~ \\
\hline
$\mu_r=\sqrt{\hat{s}}$ & 177 & 334 & 6.01 & 32.1 & 0.70 & 1.23 \\
\hline
$\mu_r=\sqrt{\hat{s}}/2$ & 205 & 385 & 6.73 & 35.9 & 0.78 & 1.37 \\
\hline
$\mu_r=M_t$ & 278 & 537 & 8.27 & 43.0 & 0.90 & 1.60\\
\hline
\end{tabular}
\caption{Total cross sections (in unit: fb) for the heavy quarkonium photoproduction under the improved conventional renormalization scale setting for three scale choices $\mu_r=\sqrt{\hat{s}}$, $\sqrt{\hat{s}}/2$, and $M_t$. $\sqrt{S}=500$ GeV. } \label{imprQ}
\end{table}

The renormalization scale in the process provides another important source of theoretical uncertainty. Under the conventional scale setting, in addition to the choice of $\mu_r=M_t$, we take other two frequently adopted choices $\mu_r=\sqrt{\hat{s}}$ and $\sqrt{\hat{s}}/2$ to do our discussion on the scale uncertainties. The results are presented in Table~\ref{convQ}. From Table~\ref{convQ}, one can see that large uncertainties $\sim62\%$ for charmonium, $\sim51\%$ for $(c\bar{b})$-quarkonium, and $\sim48\%$ for bottomonium can be obtained under three different choices of $\mu_r$, i.e. $\mu_r =M_t$, $\sqrt{\hat{s}}/2$, and $\sqrt{\hat{s}}$. The optimal renormalization scale could be determined if we have known the $\{\beta_i\}$-terms of the pQCD series~\cite{pmc}. For our present leading-order estimation, we have no $\{\beta_i\}$-terms to determine the scale. In order to minimize the conventional scale uncertainties, we adopt the improved conventional scale setting proposed in Ref.~\cite{pmc1} to do the calculation. Under such method, the next-to-leading order terms for the $\alpha_s$ running is included as a compensation for analyzing the scale errors, i.e., we substitute the following formulae into the expressions:
\begin{equation}
\alpha_s(M_t) = \alpha_s(\mu_r) \left[1 - \alpha_s(\mu_r)\frac{\beta_0}{4 \pi} \ln\left(\frac{M_t^2}{\mu_r^2} \right) \right].
\end{equation}
Numerical results are put in Table~\ref{imprQ}, in which the above three typical scales are adopted. Table~\ref{imprQ} shows that the scale uncertainties are reduced to $\sim37\%$ for charmonium, $\sim26\%$ for $(c\bar{b})$-quarkonium, and $\sim23\%$ for bottomonium.

\section{Summary}

The photoproduction of heavy quarkonium in the future $e^+e^-$ collider ILC has been studied within the NRQCD framework. To improve the calculation efficiency, the improved helicity amplitude approach has been adopted in the calculation. Total and differential photoproduction cross sections, together with their uncertainties, have been presented. The quarkonium $p_t$ distributions drop down logarithmically in the large $p_t$ region, and there is a plateau within $|y|\lesssim 4$ for the charmonium photoproduction, $|y|\lesssim 3.5$ for the $B_c$ mesons photoproduction, and $|y|\lesssim 3$ for the bottomonium photoproduction. By taking $m_c=1.50\pm0.10$ GeV and $m_b=4.9\pm0.20$ GeV, we shall have $(2.8^{+1.0}_{-0.7})\times 10^{6}$ $\eta_c$, $(5.4^{+1.9}_{-1.3})\times 10^{6}$ $J/\psi$, $(8.3^{+2.2}_{-1.8})\times 10^{4}$ $B_c$, $(4.3^{+1.1}_{-0.9})\times 10^{5}$ $B_c^*$, $(9.0^{+1.7}_{-1.4})\times 10^{3}$ $\eta_b$, and $(1.6\pm0.3)\times 10^{4}$ $\Upsilon$ events to be generated in one operation year at the ILC under the condition of $\sqrt{S}=500$ GeV and ${\cal L}\simeq 10^{36}$cm$^{-2}$s$^{-1}$. This shows that sizable amount of heavy quarkonium events can be produced via the photoproduction channels at the ILC. Thus, in addition to the hadronic experiments, the ILC shall also provide a useful platform for studying the heavy quarkonium properties.

In the present paper, we have concentrated on the dominant color-singlet mechanism via the subprocess $\gamma\gamma \to |[Q\bar{Q'}]_{\bf 1}(n)\rangle+Q'+\bar{Q}$. Within the NRQCD framework, the color-octet mechanism may also provide sizable contributions. Despite many successes of the NRQCD factorization formalism, some problems still remain unsolved. Among them a crucial one is that the approach fails to predict the polarization of $J/\psi$ and $\psi'$ at the large $p_t$ region measured at Tevatron. Thus it is helpful to find other platforms to test the NRQCD theory, such as a recent analysis of the polarized $J/\psi$ photoproduction has been done at the DESY HERA~\cite{hera}. Due to sizable amount of $J/\psi$ events can be generated at the ILC, one may predict the ILC can also be helpful for testing the color-octet mechanisms.  \\

{\bf Acknowledgement:} This work was supported in part by the Fundamental Research Funds for the Central Universities under Grant No.CQDXWL-2012-Z002, by Natural Science Foundation of China under Grant No.11275280, and by the Program for New Century Excellent Talents in University under Grant No.NCET-10-0882.

\appendix

\section{Basic elements for the helicity amplitude approach}

For self-consistency, we present some basic definitions and simplifications for the spinor product and the inner product under the helicity amplitude approach. Detailed ones can be found in Ref.~\cite{helicity1}.

In the Weyl representation, the notations $k_{\pm}$ and $k_{\o}$ for a light-like momentum $k^{\mu}$ are defined as follows,
\begin{eqnarray}
k_{\pm} &=& k_0\pm k_z, \; k_{\perp} = k_x+ik_y = |k_{\perp}|e^{i\varphi k}=\sqrt{k_+ k_-}e^{i\varphi k}\nonumber\\
\end{eqnarray}
By choosing the suitable phase, the Weyl spinors can be written as
\begin{eqnarray}
|k_+\rangle = \left(
          \begin{array}{c}
            \sqrt{k_+} \\
            \sqrt{k_+}e^{i\varphi k} \\
            0 \\
            0 \\
          \end{array}
        \right),\;\;
|k_-\rangle = \left(
          \begin{array}{c}
            0 \\
            0 \\
            \sqrt{k_+}e^{-i\varphi k} \\
            -\sqrt{k_+} \\
          \end{array}
        \right).
\end{eqnarray}
Then the basic elements in our calculation can be formulated as follow,
\begin{eqnarray}
\langle k_1 \cdot k_2 \rangle &=& \langle k_{1-}|k_{2+}\rangle \nonumber\\
&=& \sqrt{k_{1-}k_{2+}}e^{i\varphi_1}- \sqrt{k_{1+}k_{2-}}e^{i\varphi_2} \nonumber\\
&=& k_{1\perp}\sqrt{\frac{k_{2+}}{k_{1+}}} -k_{2\perp}\sqrt{\frac{k_{1+}}{k_{2+}}}.
\end{eqnarray}
For the spinor product, we have
\begin{widetext}
\begin{eqnarray}
\langle k_{1+}|{\not\!k}_3| k_{2+} \rangle = \langle k_{1+}|k_{3-}\rangle\langle k_{3-}|k_{2+}\rangle=\frac{1}{\sqrt{k_{1+}k_{2+}}}(k_{1+}k_{2+}k_{3-}-k_{1+}k_{2\perp}k_{3\perp}^*-k_{1\perp}^*k_{2+}k_{3\perp}+k_{1\perp}^*k_{2\perp}k_{3+}),
\end{eqnarray}
and for the spinor product involving the polarization vector of $^3S_1$ state with momentum $P$,
\begin{eqnarray}
\langle k_{1+}|{\not\!\epsilon}(s_z){\not\!P}| k_{2+} \rangle =&&P'_+ \epsilon(s_z)_-q_{0\perp}^*-(P'_+)^* q_{0+} \epsilon(s_z)_- - P'_+ \epsilon(s_z)_{\perp} \frac{q_{0\perp}^2}{q_{0+}}+\epsilon(s_z)_{\perp} (P'_{\perp})^* q_{0\perp}^* \nonumber\\
&& -P'_{\perp}\epsilon(s_z)^*_{\perp}q_{0\perp}^*+q_{0+}\epsilon(s_z)^*_{\perp}P'_- +\epsilon(s_z)_+ P'_{\perp}\frac{q_{0\perp}^2}{q_{0+}}+\epsilon(s_z)_+P'_{-} q^*_{0+},
\end{eqnarray}
where $P'=P-\frac{P^2}{2P \cdot q_0}q_0$. The polarization vector of the $^3S_1$ bound state with momentum $P$ ($P^2=M^2$) can be written as
\begin{eqnarray}
\epsilon^x(P) &=& \frac{|P_z|}{P_z M}\left(\frac{P_x P_0}{\sqrt{P_0^2-P_y^2-P_z^2}},\sqrt{P_0^2-P_y^2-P_z^2},\frac{P_x P_y}{\sqrt{P_0^2-P_x^2-P_z^2}},\frac{P_x P_z}{\sqrt{P_0^2-P_y^2-P_z^2}}, \right),\nonumber\\
\epsilon^y(P) &=& \left(\frac{P_y P_0}{\sqrt{P_0^2-P_z^2}\sqrt{P_0^2-P_y^2-P_z^2}},0,\frac{\sqrt{P_0^2-P_z^2}}{\sqrt{P_0^2-P_x^2-P_z^2}},\frac{P_y P_z}{\sqrt{P_0^2-P_z^2}\sqrt{P_0^2-P_y^2-P_z^2}}, \right),\nonumber\\
\epsilon^z(P) &=& \frac{1}{\sqrt{P_0^2-P_z^2}}(P_z,0,0,P_0).
\end{eqnarray}
which satisfy $\epsilon^i \cdot P=0, \;  \epsilon^i \cdot \epsilon^j=-\delta^{ij}$ ($i,j=x,y,z$).
\end{widetext}

\end{document}